\documentclass[11pt]{article}
\usepackage{amsmath}
\usepackage{amssymb}

\input{tcilatex}
\input tcilatex

\begin{document}

\author{{\small M. Rashdan}$^{1}${\small , T. Ali}$^{2}${\small , S. El-Kholy%
}$^{3}${\small , M. Abu-Shady}$^{3}$ \\
\\
{\small \ Department of Mathematics and Theoretical Physics, AEA, Cairo,
Egypt}$^{1}.$\\
{\small \ Department of Mathematics, Faculty of Science, Helwan university,
Egypt}$^{2}.$\\
{\small \ Department of Mathematics, Faculty of Science, Menoufia
University, Egypt}$^{3}$}
\title{\textbf{Hadron Properties in a Chiral Quark-Sigma Mode}l}
\maketitle

\begin{abstract}
{\ Within a chiral quark sigma model in which quarks interact via the
exchange of $\sigma $- and $\vec{\pi}$-mesons, hadron properties are
investigated. This model of the nucleon and delta is based on the idea that
strong QCD forces on very short distances (a small length scales 0.2- 1 fm)
result in hidden chiral $SU(2)\times SU(2)$ symmetry and that there is a
separation of roles between these forces which are responsible for binding
quarks in hadrons and the forces which produce absolute confinement. We have
solved the field equations in the mean field approximation for the hedgehog
baryon state with different sets of model parameters. A new parameterization
which well describe the nucleon properties has been introduced and compared
with experimental data.}
\end{abstract}

\section{INTRODUCTION}

The fundamental constituents of hadrons are quarks. Interaction of these
quarks is described by quantum chromodynamics (QCD) in terms of the exchange
of gluons. The QCD theory is a non-abelian gauge theory and has the
properties of chiral symmetry, asymptotic freedom, where the effective
coupling constant can be shown to tend to zero at short distances and
confinement, where at long distances the coupling constant in QCD\textbf{\ }%
grows. QCD Lagrangian in $u,d$ and to some extent $s$ sector is invariant
under chiral symmetry if we neglect the current masses ( m$_{u}\approx $%
2MeV,~m$_{d}\approx $5MeV,~m$_{s}\approx $150MeV) of these quarks in
comparison to the QCD scale parameter $\wedge $($\approx $ 250 MeV). It is
becoming clear now that chiral symmetry of QCD Lagrangian and its
spontaneous breaking [1] play a very important role in determining the
structure of low mass hadrons which consist of $u,d$ and $s$ quarks, and
instantaneously play a crucial role in hadron correlators in mediating the
spontaneous chiral symmetry breaking [2, 3]. Physical confinement of quarks
seems to play a lesser role. The spontaneous breaking of the chiral symmetry
is signaled by the non-vanishing values in physical vacuum of the quark and
gluon condensates [ 4-6] and we describe the solution of the mean- field
equations for the so-called hedeghog [7] baryon state. Contrary to the
claims made by other authors [7, 8] we wish to stress that the hadeghog is
not a mythical creature. This property has been interpreted as an intrinsic
wave function containing a mixture. The mean-field equations are a
straightforward extension of that outlined by Goldflam and Wilets [9] and
Birse and Banerjee [10]. Description of the model is given Sec.2. The
results and discussion are given in Sec.3.

\section{CHIRAL QUARK-SIGMA MODEL}

We describe the interactions of quarks with $\sigma -$and $\vec{\pi}$ -
mesons by Brise and Banerjee [10]. The Lagrangian density is, 
\begin{equation}
L\left( r\right) =i\overline{\Psi }\partial _{\mu }\gamma ^{\mu }\Psi +\frac{%
1}{2}\left( \partial _{\mu }\sigma \partial ^{\mu }\sigma +\partial _{\mu }%
\overrightarrow{\pi }.\partial ^{\mu }\overrightarrow{\pi }\right) +g%
\overline{\Psi }\left( \sigma +i\gamma _{5}\overrightarrow{\tau }.%
\overrightarrow{\pi }\right) \Psi -U\left( \sigma ,\overrightarrow{\pi }%
\right) ,  \tag{1}
\end{equation}

with 
\begin{equation}
U\left( \sigma ,\vec{\pi}\right) =\frac{\lambda ^{2}}{4}\left( \sigma ^{2}+%
\vec{\pi}^{2}-\nu ^{2}\right) ^{2}+m_{\pi }^{2}f_{\pi }\sigma ,  \tag{2}
\end{equation}
is the meson-meson interaction potential where the $\Psi ,\sigma $ and $\vec{%
\pi}$ are the quark , (scalar, isoscalar) sigma and (pseudoscalar,
isovector) pion fields, respectively. In the semiclassical or mean-field
approximation the meson fields are treated as time-independent classical
fields. This means that we are replacing power and products of the meson
fields by corresponding powers and products of their expectation values. The
meson-meson interactions in equ. (2) lead to hidden chiral $SU(2)\times
SU(2) $ symmetry with $\sigma \left( r\right) $ taking on a vacuum
expectation value \ \ \ \ \ \ \ 
\begin{equation}
\ \ \ \ \ \ \left\langle \sigma \right\rangle =-f_{\pi },  \tag{3}
\end{equation}
where $f_{\pi }=93$ MeV is the pion decay constant. The final \ term in equ.
(2) is included to break the chiral symmetry. It leads to partial
conservation of axial-vector isospin current (PCAC). The parameters $\lambda
^{2},\nu ^{2}$ can be expressed in terms of $\ f_{\pi }$, the masses of
mesons as, 
\begin{equation}
\lambda ^{2}=\frac{m_{\sigma }^{2}-m_{\pi }^{2}}{2f_{\pi }^{2}},  \tag{4}
\end{equation}
\begin{equation}
\nu ^{2}=f_{\pi }^{2}-\frac{m_{\pi }^{2}}{\lambda ^{2}}.  \tag{5}
\end{equation}
Now we expand the extremum, with the shifted field defined as

\begin{equation}
\sigma =\sigma ^{\prime }-f_{\pi },  \tag{6}
\end{equation}
substituting by equation(6) into equation (1) we get : 
\begin{eqnarray}
L\left( r\right) &=&i\overline{\Psi }\partial _{\mu }\gamma ^{\mu }\Psi +%
\frac{1}{2}\left( \partial _{\mu }\sigma ^{\prime }\partial ^{\mu }\sigma
^{\prime }+\partial _{\mu }\overrightarrow{\pi }.\partial ^{\mu }%
\overrightarrow{\pi }\right) -g\overline{\Psi }f_{\pi }\Psi +g\overline{\Psi 
}\sigma ^{\prime }\Psi +ig\overline{\Psi }\gamma _{5}\vec{\tau}.%
\overrightarrow{\pi }\Psi  \notag \\
&&-U\left( \sigma ^{\prime },\pi \right) ,  \TCItag{7}
\end{eqnarray}

with 
\begin{equation}
U\left( \sigma ^{\prime },\pi \right) =\frac{\lambda ^{2}}{4}\left( \left(
\sigma ^{\prime }-f_{\pi }\right) ^{2}+\vec{\pi}^{2}-\nu ^{2}\right)
^{2}+m_{\pi }^{2}f_{\pi }\left( \sigma ^{\prime }-f_{\pi }\right) ,  \tag{8}
\end{equation}

the time-independent fields $\sigma ^{^{\prime }}\left( r\right) $and $\vec{%
\pi}\left( r\right) $ are to satisfy the Euler-Lagrangian equations, and the
quark wave function satisfies the Dirac eigenvalue equation. Substituting by
equation (7) in Euler-Lagrangian equation we get:

\begin{equation}
\square \sigma ^{\prime }=g\overline{\Psi }\Psi -\lambda ^{2}\left( \left(
\sigma ^{\prime }-f\right) ^{2}+\vec{\pi}^{2}-\nu ^{2}\right) \left( \sigma
^{\prime }-f\right) -m_{\pi }^{2}f_{\pi },  \tag{9}
\end{equation}

\begin{equation}
\square \vec{\pi}=ig\overline{\Psi }\gamma _{5}\vec{\tau}\Psi -\lambda
^{2}\left( \left( \sigma ^{\prime }-f\right) ^{2}+\vec{\pi}^{2}-\nu
^{2}\right) \vec{\pi}..  \tag{10}
\end{equation}
where $\vec{\tau}$ refers to Pauli isospin-matrices, $\gamma _{5}=\left( 
\begin{array}{cc}
0 & 1 \\ 
1 & 0%
\end{array}
\right) .$

By using Hedgehog Ansatz [10] where 
\begin{equation}
\vec{\pi}\left( r\right) =\overset{\symbol{94}}{r}\pi \left( r\right) . 
\tag{11}
\end{equation}

The chiral Dirac equation for the quarks are 
\begin{equation}
\frac{du}{dr}=-p\left( r\right) u+\left( E+m_{q}-S(r)\right) w,  \tag{12}
\end{equation}
where $S(r)=g\left\langle \sigma ^{\prime }\right\rangle ,P(r)=\left\langle 
\vec{\pi}.\hat{r}\right\rangle ,E$ \ are the scalar potential, the
pseudoscalar potential and the eigenvalue of the quarks spinor $\Psi $ 
\begin{equation}
\frac{dw}{dr}=-\left( E-m_{q}+S(r)\right) u-\left( \frac{2}{r}-p\left(
r\right) \right) w,  \tag{13}
\end{equation}
The set of equations( 9,10,12,13) is solved following the method used by
Goldflam and Wilets [9] and Birse and Banarje $\left[ 10\right] $ for the
Soliton Bag model [13]. Including the color degree of freedom, one has $g%
\overline{\Psi }\Psi \rightarrow N_{c}g\overline{\Psi }\Psi $ where $N_{c}=3$
colors, $g\ \ $is coupling constant. Thus

\begin{equation}
\Psi _{{}}\left( r\right) =\frac{1}{\sqrt{4\pi }}\left[ 
\begin{array}{c}
u\left( r\right) \\ 
\\ 
iw\left( r\right)%
\end{array}
\right] \qquad and\qquad \overline{\Psi }\left( r\right) =\frac{1}{\sqrt{%
4\pi }}\left[ 
\begin{array}{cc}
u\left( r\right) & iw\left( r\right)%
\end{array}
\right]  \tag{14}
\end{equation}

then

\begin{eqnarray}
\rho _{s} &=&N_{c}g\overline{\Psi }\Psi =\frac{3g}{4\pi }\left(
u^{2}-w^{2}\right) ,  \TCItag{15} \\
\rho _{p} &=&iN_{c}g\overline{\Psi }\gamma _{5}\vec{\tau}\Psi =\frac{3}{4\pi 
}g\left( -2uw\right) ,  \TCItag{16} \\
\rho _{v} &=&\frac{3g}{4\pi }\left( u^{2}+w^{2}\right) ,  \TCItag{17}
\end{eqnarray}

where $\rho _{s},$ $\rho _{p}$ and $\rho _{v}$ are the sigma density, pion
density, and vector density respectively. These equations are subject to the
boundary conditions that asymptotically the fields approach their vacuum
values, 
\begin{equation}
\sigma \left( r\right) \sim -f_{\pi }\text{ MeV}\qquad \pi \left( r\right)
\sim 0\text{ at }r\rightarrow \infty .  \tag{18}
\end{equation}
Finally, we have solved the equations (12, 13) using fourth order
Rung-Kutta. Due to the implicit nonlinearly of our equations (9, 10) it is
necessary to iterate the solution until self-consisteny is achieved. To
start this iteration process we use the chiral circle form for the meson
fields:

\begin{equation}
S(r)=m_{q}(1-\cos \theta )  \tag{19}
\end{equation}

\begin{equation}
P(r)=-m_{q}\sin \theta  \tag{20}
\end{equation}
where 
\begin{equation}
\theta =\pi \tanh r  \tag{21}
\end{equation}

\section{\textbf{\ Numerical Results and Discussion}}

The set of equations (9-13) are solved numerically by iteration as [10] for
different values of the sigma mass, quarks mass, quark-meson coupling
constant and pion decay coupling constant. The dependence of the nucleon
properties on the sigma and quark masses and pion decay coupling constant
are listed in tables(1), (2), (3) and (4). As seen from table(2) increasing
the sigma mass the quark eigenvalue decreases, while the quark and pion
kinetic and interaction energies increase. The experimental value of the
pion decay coupling constant is of the order 93 MeV. Decreasing this value
decreases more the quark eigenvalue and slowly decrease the quark, sigma and
pion kinetic energies, while the quark-sigma and pion interaction energies
increase, as seen from table(1). The constituent quark mass has a wider
range $\sim $ 313- 470 MeV, depending on the model parameters. In fact,
there no experiment value of this parameter since the quark can not be
isolated. Birse and Banerjee [10] suggested a value 500 MeV which is very
large. Decreasing this value has the effect to increase the quark
eigenvalue, while strongly decreases the quark, sigma and pion kinetic
energies, as seen from table(3). In addition, the quark-sigma and pion
interaction energies decrease. Recently, topological and fractal models have
been introduced to calculate the constituent quark mass [14-20]. A value 336
MeV has been introduced by these models [21]. However, using this value in
the present sigma model did not lead to any convergence in the iteration
procedure for solving the field equations. We thus suggest a value of the
quark-meson coupling constant 92.3 MeV, which lead to a value 461.5 MeV for
the constituent quark mass$\left( M_{q}=gf_{\pi }\right) $ [22]. This value
is very close to that derived from NJL model which is of the order 463 MeV
[22]. The sigma mass is taken to be 900 MeV which is about twice the value
of the quark mass, as derived from NJL model. The pion mass is taken from
experiment which is of the order 139.6 MeV. This set of masses and coupling
constant predicted the correct nucleon mass (after subtracting the CM
correction), where we get a value 939 MeV, as seen from table (4) this table
shows also that the experimental magnetic moments of the proton and neutron
are well reproduced. The Goldberger-Trieman pion-nucleon coupling constant
which is of the order 1.4 is also very well reproduced. 

\textbf{Table(1}) Details of energy calculations and magnetic moments for\ $%
m_{q}=500$ MeV$,m_{\sigma }=1200$ MeV $m_{\pi }=139.6$ MeV at different
value of $f_{\pi }.$ All quantities in MeV.

\begin{tabular}{|l|l|l|l|l|l|l|l|l|}
\hline
$f_{\pi }$ & {\small 93} & {\small 91.77} & {\small 91} & {\small 90} & 
{\small 89} & {\small 88} & {\small 87} & {\small [10]} \\ \hline
{\small Quark eigenvalue} & {\small 31.995} & {\small 24.11} & {\small 19.20}
& {\small 12.88} & {\small 6.59} & {\small 3.45} & {\small -5.83} & {\small %
30.5} \\ \hline
{\small Quark kinetic energy} & {\small 1225.011} & {\small 1220.99} & 
{\small 1218} & {\small 1214.71} & {\small 1210} & {\small 1206} & {\small %
1202} & {\small 1219} \\ \hline
{\small Sigma kinetic energy} & {\small 359.66} & {\small 357.67} & {\small %
356.31} & {\small 354.4} & {\small 352} & {\small 350} & {\small 347} & 
{\small 358} \\ \hline
{\small Pion kinetic energy} & {\small 562.92} & {\small 560.98} & {\small %
559.61} & {\small 557.6} & {\small 555} & {\small 553} & {\small 550} & 
{\small 565} \\ \hline
{\small Sigma interaction energy} & {\small -183} & {\small -193} & {\small %
-198} & {\small -205} & {\small -212} & {\small -219} & {\small -226} & 
{\small -184} \\ \hline
{\small pion interaction energy} & {\small -945} & {\small -955} & {\small %
-961} & {\small -970} & {\small -978} & {\small -989} & {\small -994} & 
{\small -943} \\ \hline
{\small Meson interaction energy} & {\small 100.5361} & {\small 100.725} & 
{\small 100.899} & {\small 101.175} & {\small 101.5} & {\small 101.9} & 
{\small 102.3} & {\small 101} \\ \hline
{\small Hedgehog mass baryon} & {\small 1119.12} & {\small 1091} & {\small %
1075.45} & {\small 1051.86} & {\small 1029} & {\small 1006} & {\small 983} & 
{\small 1116} \\ \hline
{\small Total moment of proton} & {\small 2.87} & {\small 2.90} & {\small %
2.92} & {\small 2.951} & {\small 2.79} & {\small 3.} & {\small 3.02} & 
{\small (2.79)Exp.} \\ \hline
{\small Total moment of neutron} & {\small -2.29} & {\small -2.32} & {\small %
-2.33} & {\small -2.355} & {\small -2.37} & {\small -2.39} & {\small -2.41}
& {\small (-1.91)Exp.} \\ \hline
{\small Total }$g_{A}\left( 0\right) $ & {\small 1.527} & {\small 1.553} & 
{\small 1.56} & {\small 1.589} & {\small 1.85} & {\small 1.854} & {\small %
1.84} & {\small (1.25)Exp.} \\ \hline
{\small Total }$g_{\pi NN}\frac{m_{\pi }}{2M_{B}}$ & {\small 1.8606} & 
{\small 1.859} & {\small 1.857} & {\small 1.856} & {\small 1.61} & {\small %
1.63} & {\small 1.65} & {\small (1.00)Exp.} \\ \hline
\end{tabular}

\bigskip \newpage

\textbf{Table(2}) Details of energy calculations and magnetic moments for\ $%
m_{q}=500$MeV$,m_{\pi }=139.6$MeV at different values of $m_{\sigma }$
parameter. All quantities in MeV.

\begin{tabular}{|l|l|l|l|l|l|l|l|}
\hline
$m_{\sigma }$ & {\small 1200} & {\small 1100} & {\small 1000} & {\small 900}
& {\small 800} & {\small 500} & $\left[ 10\right] $ \\ \hline
{\small Quark eigenvalue} & {\small 31.995} & {\small 33.26} & {\small 34.96}
& {\small 37.26} & {\small 40.411} & {\small 58.09} & {\small 30.5} \\ \hline
{\small Quark kinetic energy} & {\small 1225.01} & {\small 1221.97} & 
{\small 1217.36} & {\small 1210.54} & {\small 1200.72} & {\small 1137.96} & 
{\small 1219} \\ \hline
{\small Sigma kinetic energy} & {\small 359.66} & {\small 360.33} & {\small %
360.65} & {\small 360.48} & {\small 359.55} & {\small 348.51} & {\small 358}
\\ \hline
{\small Pion kinetic energy} & {\small 562.92} & {\small 553.7} & {\small %
542.61} & {\small 528.86} & {\small 511} & {\small 428.23} & {\small 565} \\ 
\hline
{\small Sigma interaction energy} & {\small -183} & {\small -182} & {\small %
-180} & {\small -178} & {\small -174} & {\small -148} & {\small -184} \\ 
\hline
{\small pion interaction energy} & {\small -945} & {\small -939} & {\small %
-931} & {\small -920} & {\small -905} & {\small -815} & {\small -943} \\ 
\hline
{\small Meson interaction energy} & {\small 100.536} & {\small 102.04} & 
{\small 103.81} & {\small 105.91} & {\small 108.35} & {\small 117} & {\small %
101} \\ \hline
{\small Hedgehog mass baryon} & {\small 1119.12} & {\small 1115.95} & 
{\small 1111.99} & {\small 1107.07} & {\small 1100.83} & {\small 831.80} & 
{\small 1116} \\ \hline
{\small Total moment of proton} & {\small 2.876} & {\small 2.875} & {\small %
2.874} & {\small 2.872} & {\small 2.871} & {\small 2.87} & ({\small 2.79)
Exp.} \\ \hline
{\small Total moment of neutron} & {\small -2.29} & {\small -2.289} & 
{\small -2.282} & {\small -2.273} & {\small -2.264} & {\small -2.22} & (%
{\small -1.91) Exp.} \\ \hline
{\small Total }$g_{A}\left( 0\right) $ & {\small 1.8606} & {\small 1.85} & 
{\small 1.84} & {\small 1.83} & {\small 1.82} & {\small 1.78} & ({\small %
1.25) Exp.} \\ \hline
{\small Total }$g_{\pi NN}\frac{m_{\pi }}{2M_{B}}$ & {\small 1.5276} & 
{\small 1.5278} & {\small 1.52789} & {\small 1.528} & {\small 1.5289} & 
{\small 1.53} & ({\small 1.00) Exp.} \\ \hline
\end{tabular}

\bigskip

\textbf{Table(3}) Details of energy calculations and magnetic moments for\ $%
m_{\sigma }=1200$ MeV$,m_{\pi }=139.6$ MeV at different values of $m_{q}.$
All quantities MeV.

\begin{tabular}[t]{|l|l|l|l|l|l|l|l|}
\hline
${\small m}_{q}$ & {\small 500} & {\small 490} & {\small 480} & {\small 470}
& {\small 460} & {\small 450} & {\small [10]} \\ \hline
{\small Quark eigenvalue} & {\small 31..995} & {\small 44.77} & {\small 57.58%
} & {\small 70.47} & {\small 83.46} & {\small 96.58} & {\small 30.5} \\ 
\hline
{\small Quark kinetic energy} & {\small 1225.011} & {\small 1209.51} & 
{\small 1193.33} & {\small 1176.42} & {\small 1158.6} & {\small 1139.71} & 
{\small 1219} \\ \hline
{\small Sigma kinetic energy} & {\small 359.66} & {\small 353.8} & {\small %
347.53} & {\small 340.58} & {\small 332.97} & {\small 289.74} & {\small 358}
\\ \hline
{\small Pion kinetic energy} & {\small 562.92} & {\small 553.86} & {\small %
543.71} & {\small 533.48} & {\small 522.75} & {\small 511.44} & {\small 565}
\\ \hline
{\small Sigma interaction energy} & {\small -183} & {\small -164} & {\small %
-144} & {\small -123.35} & {\small -101} & {\small --77} & {\small -184} \\ 
\hline
{\small pion interaction energy} & {\small -945} & {\small -910} & {\small %
-876} & {\small -841.66} & {\small -806} & {\small -772} & {\small -943} \\ 
\hline
{\small Meson interaction energy} & {\small 100.5361} & {\small 100.057} & 
{\small 99.6822} & {\small 99.411} & {\small 99.244} & {\small 99.20} & 
{\small 101} \\ \hline
{\small Hedgehog mass baryon} & {\small 119.12} & {\small 1141.74} & {\small %
1163.67} & {\small 1184.91} & {\small 1205.36} & {\small 1224} & {\small 1116%
} \\ \hline
{\small Total moment of proton} & {\small 2.876} & {\small 2.867} & {\small %
2.85} & {\small 2.83} & {\small 2.824} & {\small 2.80} & ({\small 2.79) Exp.}
\\ \hline
{\small Total moment of neutron} & -{\small 2.29} & {\small -2.28} & {\small %
-2.26} & {\small -2.25} & {\small -2.23} & {\small -2.21} & ({\small -1.91)
Exp.} \\ \hline
{\small Total }$g_{A}\left( 0\right) $ & {\small 1.8606} & {\small 1.857} & 
{\small 1.854} & {\small 1.850} & {\small 1.84} & {\small 1.83} & ({\small %
1.25) Exp.} \\ \hline
{\small Total }$g_{\pi NN}\frac{m_{\pi }}{2M_{B}}$ & {\small 1.5276} & 
{\small 1.51} & {\small 1.50} & {\small 1.48} & {\small 1.46} & {\small 1.45}
& ({\small 1.00) Exp.} \\ \hline
\end{tabular}

\bigskip \textbf{Table(4}) Details of energy calculations and magnetic
moments for $m_{q}=461$MeV$,m_{\pi }=139.6$MeV$,m_{\sigma }=900$MeV$,$ $%
f_{\pi }=92.3\,$MeV$.$ All quantities MeV.

\begin{tabular}{|l|l|}
\hline
{\small Quark kinetic energy} & {\small 1134.58} \\ \hline
{\small Sigma kinetic energy} & {\small 331.9} \\ \hline
{\small Pion kinetic energy} & {\small 477.09} \\ \hline
{\small Sigma interaction energy} & {\small -98} \\ \hline
{\small Pion interaction energy} & {\small -778} \\ \hline
{\small Total without C. M. correction} & {\small 939.3} \\ \hline
{\small Meson interaction energy} & {\small 106.62} \\ \hline
{\small Hedgehog mass baryon} & {\small 1172} \\ \hline
{\small Total moment of proton} & {\small 2.82} \\ \hline
{\small Total moment of neutron} & {\small -2.22} \\ \hline
{\small Total }$g_{A}\left( 0\right) $ & {\small 1.822} \\ \hline
{\small Total }$g_{\pi NN}\frac{m_{\pi }}{2M_{B}}$ & {\small 1.484} \\ \hline
\end{tabular}

\section{References}

\begin{enumerate}
\item V. Koch, Int. J. Mod. phys. E6, 203 (1997) and references therein.

\item T. Schafer and Evshuryak, Rev. Mod. Phys. 70, 323 (1998).

\item D. Diakanov, hep-ph/9602375, talk given at International School of
Physics, $^{\prime }$Enrico Fermi$^{\prime }$,

course 80 (Varenna,Italy,1995)

\item Y. Nambu and G. Jona-Lasinio, Phys. Rev. 122, 345 (1961); 124(1961).

\item M. Gell-Mann, R. Joakes and B. Renner, Phys. Rev. 175, 2195 (1968).

\item M. A. Shifman, A. I. Vainshtein and V. I. Zakharov, Nucl. Phys. A511,
676 (1990).

\item A. Chodos and C. B. Thorn, Phys. Rev. D12, 2733 (1975).

\item V. Vento, J. G. Jun, E. M. Nyman, M. Rho, and G. E. Brown Nucl. Phys.
A345, 413(1980).

\item R. Goldflam and L. Wilets, Phys. Rev. D25, 1951 (1982).

\item M. Birse and M. Banerjee, Phys. Rev. D31, 118 (1985).

\item J. Rakelski, Phys. Rev. D16, 1890 (1977).

\item R. F. Dashen, B. Hasslacher, and A. Neveu, Phys. Rev. D 10 ,
4114(1974).

\item R. Friedberg and T. D. Lee, Phys. Phys. Rev. D15, 1694 (1977), 16,
1096 (1977).

\item M.S. El Naschie, Chaos, Solitons Fractals, 2, 211 (1992)

\item B. Mandelbrot, 1983, The fractal geometry of nature (Freeman).

\item L. Nottale, Int. J. Mod. Phys. A7, 4899(1992).

\item L. Nottale, Chaos, Solitons fractals 7, 877 (1966); AA lett. 315, L(
(1996);

AA 327,867 (1997)

\item Nottale, Chaos, Solitons Fractals 9, 1035 (1998); Chaos, Solitons
Fractals 9, 1043

\item G. N. Ord, Chaos, Solitons Fractals 7, 821 (1992)

\item S. Weinberg Rev. Mod. Phys. 61, 1 (1989)

\item M.S. El Naschie, Chaos, Solitons Fractals, 14, 369 (2002)

\item M. Rashdan, Chaos, Solitons Fractals, 18, 107 (2003).
\end{enumerate}

\end{document}